\documentclass{conm-p-l}

\newtheorem{theorem}{Theorem}[section]

\theoremstyle{definition}
\newtheorem{definition}[theorem]{Definition}

\begin{document}

\title{Opaque predicates, veiled sets and their logic}
\thanks{This paper was
written when D\'ecio Krause was a Visiting Scholar in the School of
Philosophy, University of Leeds, partially supported by CNPq (Philosophy
Section). The authors would like very much to thank the two anonymous
referees for their suggestions.}

\author{D\'ecio Krause}
\address{Department of Mathematics,
Federal University of Paran\'a, P.O.Box 19081, Curitiba, PR, Brazil}
\email{dkrause@mat.ufpr.br}
 
\author{Steven French}
\address{Division of History and Philosophy of Science,
School of Philosophy,
University of Leeds, Leeds LS2 9JT, UK}
\email{s.r.d.french@leeds.ac.uk}

\subjclass{Primary 03B60; Secondary 03B15}
\date{}

\begin{abstract}
Motivated by considerations in the foundations of quantum
mechanics and inspired by the literature on vague predicates, we
introduce the concept of an {\it opaque predicate\/}.  While in
the case of vague predicates there is a kind of indeterminacy
with respect to the predicate, in the sense that the vagueness
concerns whether a well-determined object satisfies it or not,
in the case of opaque predicates the indeterminacy is with
regard to the objects which should satisfy them. In other words,
their extensions are not well-defined, despite the fact that the
conditions for an object to satisfy the predicates are
well-known.  We suggest that such opaque predicates (and more
generally, what we call {\it opaque relations\/}) can be
characterized by a logic which encompasses a semantics founded
in quasi-set theory, and call their extensions {\it veiled
sets\/}.  
\end{abstract}

\maketitle

\section{Vagueness and Opacity}

\makebox[2.0in]{}
\parbox{2.5in}
{
\begin{small}
{\sf
``Vagueness is a feature of scientific as of other discourse.''

\vspace{2mm}
\hfill{Max Black (1966)}}
\end{small}}

\vspace*{4mm}
Peirce famously characterised vagueness in the following terms:
``A proposition is vague when there are possible states of things
concerning which it is intrinsically uncertain whether, had they
been contemplated by the speaker, he would have regarded them as
excluded or allowed by the proposition. By intrinsically
uncertain we mean not uncertain in consequence of any ignorance
of the interpreter, but because the speaker's habits of language
were indeterminate'' \cite{pei02}. 

In this context, two
issues immediately arise: (i) whether vagueness can be dismissed
as merely a feature of `natural' language which will effectively
evaporate with the introduction of some formal system, and (ii) given
a negative answer to (i), whether vagueness requires the use of
some form of non-classical logic.  These are the issues with
which standard discussions of vagueness have been concerned.
Thus Wright, for example, emphasises that vague predicates lack
`sharp boundaries' and argues forcefully that ``... the utility
and point of the classifications expressed by many vague
predicates would be frustrated if they were supplied with sharp
boundaries'' \cite[p. 227]{wri76}. In his own terms, the most
`profound' example he gives is that of colour predicates, where
the elimination of vagueness would incur the price of
jeopardising contact between language and empirical reality.
Thus, vagueness is a phenomenon of `semantic depth', in the
sense that ``[i]t is not usually a matter simply of our lacking
an instruction where to `draw the line'; rather the instructions
we already have determine that the line is not to be drawn''
[ibid.]. However, if `the line is not to be drawn', then clearly
classical logic is not to be used. Following Putnam \cite{put83}, Mott
has recently suggested that vague predicates can be accommodated
within intuitionistic logic if they are taken to be partial, in
precisely the above sense; that is, in certain situations, their
application simply cannot be decided at all \cite{mot94}. Thus,
the application of a colour predicate can be decided simply by
using our sense organs, in those situations where it can be
decided at all and it is precisely because of the existence of
the complement of such situations that the predicate can be
termed `vague'. At the end of his paper Mott suggests that
intuitionistic mathematics might be the appropriate framework
for quantum physics, noting that ``[i]f sometimes there really is
no fact of the matter whether a certain object is a table, or a
certain shade is red, or a certain man bald, then perhaps
sometimes there is no fact of the matter exactly where an
electron is either'' [ibid., p. 147].  

However, to talk of there
being no fact of the matter as to the location of an electron is
to assume that one is talking about a well-defined individual,
such that the application of the predicate regarding its
location cannot be determined. As we have indicated elsewhere, this is
certainly one way of considering the quantum situation (\cite{low94},
\cite{frekra95}, \cite{frekramai95}).
Nevertheless, our approach here is different and has not to our
knowledge been explored in the literature. Quantum particles may
also be regarded as `non-individuals' in a certain sense \cite{pos63}
(see \cite{fre89}, \cite{krafre95})
 and now the issues shifts from the applicability of the
predicates to the determinateness of the objects. It is this
which marks the difference between vagueness and what we shall
call opacity. With
regard to the latter, it is not an issue of the predicate
lacking `sharp boundaries' but rather of the objects to which
the predicate applies lacking individuality.  In the case of
opaqueness, then, the grounds for dismissing it as an aspect of
natural language are even weaker than the case of vagueness. We
would like to emphasise this point: if what quantum mechanics
tells us about how the world could be\footnote{We have chosen
this formulation as neutral between the claims of realist and
anti-realist; see \cite{van91}, for example.}
is taken seriously,
as we think it should be, then, in the sense delineated here,
this `world' is opaque (cf. M. Black who argues for an analysis of
vagueness in order to avoid the ``wholesale destruction of the
formal sciences'' \cite[p. 27]{bla66}). Granted this, a completely different
formal framework is required to capture this opacity --one which
includes both syntactic and semantic elements. It is precisely
such a framework that we  sketch below.  

From a more mathematical point of view,  
vague predicates are considered to  differ
 from the usual `Fregean' predicates  (\cite{tertri89}) in the
following sense. A unary predicate letter $P$ of (say) a
first-order language is {\it Fregean\/} if it provides a bipartition on
a domain $D$ of objects to which the language makes reference.
In other words, there exist $D_{1}$ and $D_{2}$ such that
 $D_{1} \cap D_{2} = \emptyset$ ,
 $D_{1} \cup D_{2} = D$ and 
 $D_{1} = \{x \in D : P(x)\} \not= \emptyset$. 

The set $D_{1}$ is the extension of $P$ and if  $a \in D_{1}$, we say that 
`$a$ has the property $P$'; otherwise, that is, if
 $a \in D_{2}$, we say that `$a$ does not have the property $P$'. 
Vague predicates are
then characterized as those predicates which do not provide such a bipartition
in the domain. In other words, there are objects $a \in D$ such that
neither $a \in D_{1}$ nor $a \in D_{2}$ holds. For these objects, it 
is  `vague' whether 
they have or not have the property ascribed by $P$.

The idea of vagueness provided by such an analysis, 
we emphazise, is concerned with the vagueness of the
predicates involved, and not with the objects the language is making
reference to. In fact, if we consider if a certain
(well-determined) person,  who could be classified
as a philosopher, is or is not a profound thinker, then 
it may be vague whether he/she is profound or not, since the criteria for
`profundity' is vague. The same occurs with `tall', `intelligent'
and so on. Returning to quantum mechanics,
 there are situations which simply cannot be characterized
in terms of `vagueness' in the above sense. We prefer to
classify these cases in another fashion and call  the
predicates involved {\it opaque\/}.  
The following example, we hope, reinforces
what was said above.  Suppose we are dealing with a collection of $n$
electrons and that we intend to measure their spin 
 in a certain chosen direction. In other words, we might 
consider the predicate `to have spin up in the (chosen)
direction'.  It is known that physicists are able to specify
precisely what conditions electrons must obey in order to
satisfy the predicate, hence the situation is not one concerning
vagueness in the sense explained above. However, 
 physicists
have no means to determine {\it which\/}  are the electrons of the
aggregate that have spin up in the given direction. In fact, it
turns out that there may be  $m$ electrons ($m
< n$), say, with spin up in that direction, but if another measurement
is made, despite finding the same number $m$ of electrons with
spin up in the considered direction, 
there is no way of assuring that both collections
coincide; that is, we have no grounds for asserting that the electrons of this
last collection are `the same' as those of the former.
There is a strong indeterminacy here concerning the
objects of such collections, and this is one of the basic 
metaphysical interpretations of  quantum mechanics. 

Another example might be the following:  Suppose we are considering the
six electrons there are in the level 2p of a sodium atom. All the electrons
coincide with respect to {\it all\/} their quantum numbers, so there
is no way of distinguishing them. Even so, physicists reason as if
there are six `entities' (it is difficult to use the word `individuals'
in this case ---see \cite{tel95}) in that level. Then  consider
the predicate `to be one of the electrons of the level 2p of a sodium atom'.
How may we ascribe to this predicate a well-defined extension? This is not
possible without ambiguity. 

Situations of this sort, in which the indeterminacy resides not with
the predicate, but with the individuals instead, motivates our  
discussion on
opaque predicates. The question to be answered now  concerns 
the mathematical treatment of these entities. 


\section{A logic with opaque predicates}
 The basic intuitive idea 
of a  logical system encompassing opaque predicates is that
for such a predicates, their extensions cannot be defined as standard 
{\it sets\/} since such a set ({\it Menge\/}) is, according to the well
known `definition' coined by
Cantor, ``any collection into a whole of definite and separate objects
of our intuition or of our thought'' \cite[p. 85]{can58}. 
In accordance with what we have said above,
the goal is to represent the idea that
 it is not possible to {\it distinguish\/} (that is, to 
 `separate') in a strong sense
between the individuals that satisfy the predicate. 

Our strategy will be as follows. We intend to characterize as opaque those
predicates whose extension is a collection of objects (technically,
a {\it quasi-set\/}) whose elements cannot be distinguished from one another.
The problem is that, if we consider the predicates of  a certain language,
say a first-order language, we have no criteria for distinguishing
 among those
predicates which should be considered as opaque. The system described
below permits a clear definition of opaque predicates in accordance with
the above motivations. 
The reasons to use such a logic will be explained in the last section.

\vspace{3mm}
In what follows we present the main features of the 
 logic  ${\mathfrak L}_{op}$, which  is a slight modified version
of the system $S_{\omega}$ presented in \cite{coskra94}.
Let us begin by defining the concept of {\it type\/}:
we call $\Pi$ the set of {\it types\/},
recursively defined as being the  smallest
set such that: (a) $e_{1}, e_{2} \in \Pi$, and    
(b) if $\tau_{1}, \ldots, \tau_{n}  \in \Pi$, then $\langle
\tau_{1}, \ldots ,\tau_{n} \rangle \in \Pi$.

$e_{1}$ and $e_{2}$ are the types of the {\it individuals}; the
objects of type $e_{1}$ are called `$m$-objects' (short for `microobjects')
and may be  intuitively
 thought of as elementary particles of modern physics as in the example of
opacity we have described in the previous section. The objects of
type $e_{2}$ are such that
 classical logic applies to them in all its aspects. 

In more precise words, ${\mathfrak L}_{op}$
 is a higher-order logic whose language has the
following categories of primitive symbols: connectives: $\neg$ and $\to$ 
($\wedge$, $\vee$ and $\leftrightarrow$
are introduced as usual),
 the universal quantifier $\forall$ ($\exists$ is defined in the
standard way), parentheses and comma. 
With respect to
variables and constants, for each type $\tau \in
\Pi $ there exists a denumerably
infinite collection of variables $ X_{1}^{\tau}, X_{2}^{\tau}, \ldots$ of
type $\tau$ and a (possibly empty) set of constants $A_{1}^{\tau},
A_{2}^{\tau}, \ldots$
of that type; we use $X^{\tau}$, $Y^{\tau}$ and $C^{\tau}$, $D^{\tau}$ 
perhaps with subscripts as
metavariables for variables and constants of type $\tau$ 
respectively. 

The terms of type $\tau$ are the variables and the
constants of that type; so, we have individual terms of type
$e_{1}$ and of type $e_{2}$. We use $U^{\tau}$, $V^{\tau}$, perhaps with
subscripts, as syntactical variables for terms of type $\tau$.
The atomic formulas are defined in the usual way: if $U^{\tau}$
is a term of type $\tau = \langle \tau_{1}, \ldots, \tau_{n}
\rangle$ and $U^{\tau_{1}},
\ldots, U^{\tau_{n}}$ are terms of types $\tau_{1}, \ldots, \tau_{n}$
respectively, then $U^{\tau}(U^{\tau_{1}}, \ldots,
U^{\tau_{n}})$ is an atomic formula. Other formulas are defined as
in a standard way. 

The postulates  of  ${\mathfrak L}_{op}$ (axiom schemata and inference
rules) are the following:

\begin{description}
\item[(A1)]  $A$, where $A$ comes from a tautology in $\neg$ and $\to$ by
uniform substitution of formulas of ${\mathfrak L}_{op}$ for the variables.
\item[(A2)]  $\forall X^{\tau} (A \to B) \to (A \to \forall X^{\tau} B)$, 
where $X^{\tau}$ does not occur free in $A$.
\item[(A3)]  $\forall X^{\tau} A(X^{\tau}) \to A(U^{\tau})$ 
where $U^{\tau}$ is a term free for $X^{\tau}$ in 
$A(X^{\tau})$ and of the same type of $X^{\tau}$.
\item[(R1)]  From $A$ and $A \to B$ to infer $B$
\item[(R2)]  From $A$ to infer $\forall X^{\tau} A$
\end{description}

The syntactical concepts of free and bound occurences of a variable in 
a term or in
a formula, such as those of sentence (closed formula), theorem,
consistent set of formulas, etc.  can be defined
without difficulty. The logic ${\mathfrak L}_{op}$ still encompasses a
comprehension axiom, which can be stated as follows: 

If $U^{\tau}(X^{\tau_{1}}, \ldots, X^{\tau_{n}})$ is a formula in which
the variables $X^{\tau_{1}}, \ldots, X^{\tau_{n}}$ are free and if 
$X^{\tau}_{i}$ is
a predicate of type $\tau = \langle \tau_{1}, \ldots, \tau_{n} \rangle$,
then 

\begin{description}
\item[(A4)]
$\exists X^{\tau}_{i} \forall X^{\tau_{1}}, \ldots, \forall X^{\tau_{n}} (
X^{\tau}_{i}(X^{\tau_{1}}, \ldots, X^{\tau_{n}}) \leftrightarrow 
U^{\tau}(X^{\tau_{1}}, \ldots, X^{\tau_{n}}))$
\end{description}

The concept of identity is introduced for all objects except 
those denoted by terms of type $e_{1}$. This justifies  
the distinction we considered between the types of 
the individuals. The idea conforms itself
with Schr\"odinger's (see also \cite[pp. 27--29]{bor43}, \cite[pp.
49--51]{hes70}), 
 who said that the concept of identity does not make
sense for the elementary particles of modern physics \cite[pp. 17--18]{sch52},
\cite{coskra94}. The definition is briefly stated as follows:

\begin{definition} If $\tau \not= e_{1}$, then:
$$U^{\tau} = V^{\tau} \iff \forall X^{\langle \tau \rangle} (
X^{\langle \tau \rangle} (U^{\tau}) \leftrightarrow 
 X^{\langle \tau \rangle} (V^{\tau}))$$
\end{definition}

Then our logic characterizes also in a syntactical way a certain
category of objects (namely, those denoted by the terms of type
$e_{1}$) about which we cannot say either that they are identical or that
they are distinct. These objects play the role of those objects 
whose aggregates are intended 
to constitute the extensions of the opaque predicates. In this
sense, we can introduce in a more precise way 
the following definition of what is to
be understood by an {\it opaque relation\/}:

\begin{definition} An opaque relation is any term $U^{\tau}$ of type $\tau = 
\langle \tau_{1}, \ldots
\tau_{n} \rangle$ where every $\tau_{i}$ is obtained recursively 
from $e_{1}$ (that is, for every $i$, $\tau_{i}$ is $e_{1}$ itself, or
 $\langle e_{1} \rangle$, or $\langle \langle e_{1} \rangle \rangle$,
or $\langle e_{1}, e_{1} \rangle$ and so on). 
\end{definition}

It must be realized that we do not intend to enter here into the 
familiar discussion
on what is to be considerd as a `predicate'. We simply use a logical
terminology and call a `predicate' an unary relation of type $\langle
i \rangle$, where $i$ is $ e_{1}$ or $e_{2}$. Then, {\it opaque predicates\/}
are particular cases of the opaque relations expressed
 by the above definition.

The problem, as we will recall in the beginning of the next
section, is to find an adequate way of interpreting these
predicates, since it would make no sense to ascribe a {\it
set\/}, that is, a collection of distinguishable objects, as the
extension of such predicates.

\vspace{3mm}
We can add axioms of extensionality and infinity in the standard
way  to our logic by adapting those of \cite[Chap. 4]{hilack51}. 
Let us mention here only the case of the axiom of
choice, which might be thought as problematic due to the senselessness
 of the concept
of identity regarding the objects of type $e_{1}$. We may use 
 a weaker form which covers only those situations in which
there are no objects of type $e_{1}$ involved, which can be adapted
from that presented in \cite[p. 156]{hilack51}. More precisely, if
$X^{\tau_{1}}_{1}$ and $X^{\tau_{1}}_{2}$ are variables such 
that $\tau_{1} \not= e_{1}$, $Y^{\tau_{2}}_{1}$ and $2^{\tau_{1}}_{2}$ 
are variables such  that $\tau_{2} \not= e_{1}$, 
$Z^{\langle \tau_{1}, \tau_{2} \rangle}_{1}$ and 
$Z^{\langle \tau_{1}, \tau_{2} \rangle}_{2}$ are also variables, then
the axiom is:

\begin{multline*}
\forall Z^{\langle \tau_{1}, \tau_{2} \rangle}_{1} \exists
Z^{\langle \tau_{1}, \tau_{2} \rangle}_{2} (\forall 
X^{\tau_{1}}_{1} (\exists Y^{\tau_{2}}_{1} 
(Z^{\langle \tau_{1}, \tau_{2} \rangle}_{1}(X^{\tau_{1}}_{1}, Y^{\tau_{2}}_{1})
\to \\
 \exists Y^{\tau_{2}}_{1} (Z^{\langle \tau_{1}, \tau_{2} \rangle}_{2}
(X^{\tau_{1}}_{1},Y^{\tau_{2}}_{1}) \wedge  
Z^{\langle \tau_{1}, \tau_{2} \rangle}_{1}(X^{\tau_{1}}_{1},Y^{\tau_{2}}_{1})))
\to \\
 \forall X^{\tau_{1}}_{1} \forall X^{\tau_{1}}_{2} \forall 
Y^{\tau_{1}}_{1} \forall Y^{\tau_{1}}_{2} (
Z^{\langle \tau_{1}, \tau_{2} \rangle}_{2}(X^{\tau_{1}}_{1},
Y ^{\tau_{2}}_{1})
\wedge Z^{\langle \tau_{1}, \tau_{2} \rangle}_{2}(X^{\tau_{1}}_{1},
Y^{\tau_{2}}_{2}) \to Y ^{\tau_{2}}_{1} = Y ^{\tau_{2}}_{2}))
\end{multline*}

Other usual syntactical notions are defined in the standard way,
such as for instance the concept of $\vdash A$, 
 $\Gamma \vdash A$ for a set $\Gamma$ of sentences, and so on.  

Now let us consider the semantical counterpart of our logic.


\section{Semantics}
\subsection{The mathematical framework}
Let us recall once more that within the scope of the logic ${\mathfrak L}_{op}$,
we can consider objects for which there is no meaning in talking
 about either their identity or their diversity. 
But, when we consider the semantical aspects of such
a logic, in the sense of an association of certain objects of a
mathematical structure to the terms of the language, it is 
convenient that such a procedure should reflect the intuitive aspects
of the logic. Hence, it is convenient that the terms of type $e_{1}$
(those to which the usual concept of identity cannot be applied) 
should not have a well-defined interpretation. In fact, if we ascribe to the
 constants of the type
$e_{1}$ well-determined individuals of the domain (elements of a set,
say, as in the usual semantics), since 
to the elements of a set the relation of equality makes sense, 
 we are leaving aside the basic idea we intend to capture,
namely, that of the `non-individuality' of certain elements. 
So, a `natural' semantics
for our logic should be presented by using a mathematical device in which
we can talk about objects that cannot be individualized, that is,  
strongly indistinguishable objects.\footnote{In classical logic and mathematics,
the concept of `indistinguishability' cannot be separated from that
of `identity' by force of Leibniz' Law. This `identification' of
identity and indistinguishability (agreement with respect to
properties), let us recall, is in the core of Ramsey's criticism
of the treatment of identity presented in {\it Principia Mathematica\/} ({\it
vis.\/}, Leibniz' Law). In a certain sense, our logics vindicate Ramsey's
claim that ``There is nothing self-contradictory ($\ldots$) in $a$ and $b$
[where $a \not= b$] having all their elementary properties in common. 
Hence, since this is
logically possible, it is essential to have a symbolism which allows us to
consider this possibility and does not exclude it by definition''
\cite[p. 182]{ram78}.}

In other words, we are not acting in conformity with the
intuitive aspects we intend to capture if we use standard set
theory to build the mathematical structure in which the language
of ${\mathfrak L}_{op}$ is to be interpreted. Thus, instead of using
a set theory like Zermelo-Fraenkel in the metamathematics, we
use the quasi-set theory ${\mathfrak Q}$ (for details, see
\cite{kra96}).\footnote{As explained in that work, there have been
presented several
versions of quasi-set theory (in reality, they constitute distinct
theories; see our references).}

Roughly speaking,
the theory ${\mathfrak Q}$ is a mathematical device for treating
col\-lec\-tion\-s of ind\-is\-tin\-guish\-able ob\-ject\-s.  The
the\-o\-ry  allows the presence of a certain kind of {\em Urelemente\/}
 (the
so called $m$-atoms) to which the usual con\-cept of identity does not apply.
The underlying logic of ${\mathfrak Q}$ is classical quantificational logic
without identity; the specific symbols are three unary predicate
letters $m(x)$ (read `$x$ is an $m$-atom'), $M(x)$ (read `$x$ is
an $M$-atom' --that is, a standard {\em Urelement\/}) and $Z(x)$ (read `$x$
is a standard set'). Such `sets' are characterized in ${\mathfrak Q}$ as
 quasi-sets whose transitive closure
do not contain $m$-atoms). Furthermore, the language still
encompasses two binary predicate symbols $\in$
(membership) and $\equiv$ (indistinguishability) and a unary
functional symbol $qc$ (quasi-cardinality). A quasi-set (qset for
short) is defined as an entity which is not an {\em Urelement\/} (that
is, it is anything that is neither an $m$-atom nor a classical
atom). We write $Q(x)$ for saying that $x$ is a quasi-set. 
 The concept of quasi-cardinal is introduced in such a
way  that it extends the concept of cardinal for arbitrary
qsets; some additional remarks on this concept shall be mentioned below.

The axioms of indistinguishability state that $\equiv$
has the properties of an equivalence relation. 
The  {\it
extensional equality} $=_{E}$ is defined in the following way: $x =_{E} y$
iff $(Q(x) \wedge Q(y) \wedge 
\forall z (z \in x \leftrightarrow z \in y)) \vee (M(x)
\wedge M(y) \wedge x \equiv y)$. That is, in 
 extensional entities are indistinguishable
standard {\em Urelemente\/} or qsets which have `exactly the
same' elements. It can be proven that the extensional equality
has all the formal properties of classical
identity.\footnote{Hence, by `indistinguishable standard {\it
Urelemente\/}', as in the last phrase, we  understand
`identical {\it Urelemente\/}'. From now on, we will use `$=$' instead
of `$=_{E}$'.} The substitutivity principle is
valid only with respect to indistinguishable objects which are
not $m$-atoms. In symbols, $\forall x \forall y (\neg m(x)
\wedge \neg m(y) \to (x \equiv y
\to (A(x,x) \to A(x,y))))$ with the usual syntactical restrictions.
Furthermore, the axioms for the concept of quasi-cardinal
generalize the concept of cardinal for arbitrary qsets. Since
the identity relation cannot be applied to (a pair of) $m$-atoms
and since it is postulated that every qset has a quasi-cardinal,
then there is a precise sense in saying that the objects of a
qset whose elements are indistinguishable $m$-atoms can only be
aggregated in certain quantities, but that they cannot be
ordered or counted.\footnote{Then the $m$-atoms have some of
the characteristics generally attributed to quanta. See \cite{tel95}.}

One of the most peculiar axioms of ${\mathfrak Q}$ is the `weak'
axiom of extensionality, which states that qsets having `the same quantity
of elements of the same sort'  are indistinguishable qsets
(this idea can be stated conveniently by means
of the concept of quasi-cardinal and by passing the quotient by
the relation of indistinguishability). 
Among other things, this axiom permits us to prove an interesting result,
which we call `the theorem of the unobservability of permutations'
\cite{krafre95,kra96}: if in a qset we exchange one $m$-atom by an
indistinguishable one, then the resulting qset is
indistinguishable from the original qset. This can be viewed as
the formal counterpart of the unobservability of particle
permutations in quantum mechanics.\footnote{The concept of `one'
$m$-atom can be made precise by using the `strong singleton' of
an $m$-atom. More precisely, the strong singleton of $x$ is a qset
which is a subqset (in the usual sense) of the
qset of all the objects indistinguishable from $x$ (this qset
is provided by an axiom like the pair axiom of ZFC, but using
the indistinguishability relation instead of  equality) which
has quasi-cardinal $1$. In ${\mathfrak Q}$, we can prove that such a qset
exists.}
Important qsets for our purposes here are the `pure' qsets, that
is, those qsets whose elements are $m$-atoms only. Certain particular
pure qsets
will be taken to be the extensions of the opaque predicates of the
logic ${\mathfrak L}_{op}$, as we will see below.  

Furthermore, it is convenient to recall that ${\mathfrak Q}$
involves standard mathematics, in the sense that all axioms
of Zermelo-Fraenkel set theory may be suitably translated into
the language of quasi-sets and (these translations) proven as
theorems of ${\mathfrak Q}$;\footnote{For further information on this
particular topic, see for instance
\cite{kra95a,coskra95f,kra96}.} so, all the
set-theoretical operations (union, cartesian products,
difference between quasi-sets, etc.) can be performed in ${\mathfrak Q}$ 
similarly as in
standard set theory; consequently, we don't need to pay attention here to
the terminology, which is used as in the standard set theories.  

Let us still mention in brief about an axiom which states that
$qc(P(x)) = 2^{qc(x)}$, where $P(x)$ is the quasi-set of
all the subquasi-sets of $x$ (defined in the usual way).
 Here, $2^{qc(x)}$ may
 be understood as follows.
If $\alpha$ is the cardinal 
which is the quasi-cardinal of $x$,\footnote{Another axiom says that
every quasi-set has an unique  quasi-cardinal which is a cardinal 
defined in the `classical' part of ${\mathfrak Q}$. So, the quasi-cardinals
are `sets' in ${\mathfrak Q}$, obeying the standard axioms of {\sf ZF} set
theory.} then $2^{qc(x)}$ is $card(^{\alpha}2)$, that is, the cardinal of
the collection (which is a `set') of all functions from $\alpha$ to
$2 = \{0, 1\}$.\footnote{The standard concept of `function' may also be
restricted in ${\mathfrak Q}$, since an usual mapping in general cannot 
distinguish between $m$-atoms which are in the relation $\equiv$. So, 
the concept of function was extended to a concept of {\it quasi-function\/},
which intuitively speaking maps indistinguishable objects in 
indistinguishable objects. When these objects are `classical',  
the quasi-functions turn out to be `functions' in the standard sense.
As remarked, all the details may be found in \cite{kra96}.}
 It should also be realised that since both 
$\alpha$ and $2$ are `classical' objects (in the
sense that they obey the axioms of {\sf ZF}), then $^{\alpha}2$ is also
a `set' and hence the definition of $card(^{\alpha}2)$ makes sense. 

This axiom has an important consequence. 
As suggested above, in ${\mathfrak Q}$ there may exist qsets whose
elements are $m$-atoms only, called `pure' qsets. Furthermore,
it may be the case that the $m$-atoms of a pure qset $x$ are
indistinguishable from one another, in the sense of sharing the
indistinguishability relation $\equiv$. In this case, the
axioms provide the grounds for saying that nothing in the
theory can distinguish among the elements of $x$. But, then
 one could ask what is it that sustains the idea that there
is more than one entity in $x$. The answer is obtained through
the above mentioned axioms (among others, of course). Since the
quasi-cardinal of the power qset of $x$ has quasi-cardinal
$2^{qc(x)}$, then if $qc(x) = \alpha$, it results that 
for every quasi-cardinal
$\beta \leq \alpha$ there exists a subquasi-set $y \subseteq x$
such that $qc(y) = \beta$, according to the remaining axioms about the
quasi-cardinality of the subquasi-sets. Thus, if $qc(x) = \alpha
\not= 0$, the axioms do not forbid the existence of $\alpha$ 
subquasi-sets of $x$ which can be regarded as `singletons'.

Of course the theory cannot prove that these `unitary'
subquasi-sets (supposing now that $qc(x) \geq 2$) are distinct,
since we have no way of `identifying' their elements. These `singletons'
are {\it indistinguishable\/} in the sense of the axiom of 
weak extensionality. But what is important  is that quasi-set
theory is compatible with the existence of distinct and
abosolutely indistinguishable $m$-atoms. This is important, for example, for
obtaining a description of  quantum statistics in the framework
of ${\mathfrak Q}$ (see \cite{krasanvol98}
for details on this point). In other words, it is
consistent with ${\mathfrak Q}$ to maintain that $x$ has $\alpha$
elements, which may be regarded as absolutely indistinguishable
objects. Since the elements of $x$ may share the relation
$\equiv$, they may be further understood as belonging to a same
`equivalence class' (for instance, being indistinguishable
electrons) but in such a way that we cannot assert either that
they are identical or that they are distinct from one another
(i.e., they act as `identical electrons' in the physicist's
jargon).\footnote{The application of this formalism to the
concept of non-individual quantum particles has been proposed in
\cite{krafre95}.}


\subsection{The Generalized  Semantics of ${\mathfrak L}_{op}$}
All the developments of this section are performed in the
quasi-set theory ${\mathfrak Q}$. We will not provide here all the
details but only the main definitions and results, which are
similar to those of classical logic (see for instance
\cite{chu56}). The proofs can be adapted
without difficulty from the most general case of Sch\"odinger
logics presented in \cite{dckra97i}. 

  We call ${\mathfrak L}$ the language of ${\mathfrak L}_{op}$. Let
$D$ be a quasi-set such that $D = m \cup M$ where $m$ is a pure
qset whose elements are indistinguishable from one another and
$M$ is a non-empty set (in ${\mathfrak Q}$).

By a {\bf frame} for ${\mathfrak L}$ based on $D$ 
we mean a quasi-function $\mathfrak M$
whose domain in the set $\Pi$ of types such that:

\begin{enumerate}
\item ${\mathfrak M}(e_{1}) = m$ 
\item ${\mathfrak M}(e_{2}) = M$
\item For each type $\tau = \langle \tau_{1}, \ldots ,\tau_{n}
\rangle \in \Pi$, ${\mathfrak M}(k) \subseteq {\mathfrak P}({\mathfrak
M}_{{\tau}_{1}}, \times \ldots \times {\mathfrak M}_{{\tau}_{n}})$. If the
inclusion in this last expression can be replaced by (extensional) 
equality, then the frame is {\em standard\/}.
\end{enumerate}

If we write ${\mathfrak M}_{\tau}$ instead of ${\mathfrak M}(\tau)$, then the
frame can be viewed as a family (${\mathfrak M}_{\tau})_{\tau \in
\Pi}$ of qsets satisfying the above conditions. In what follows,
we will refer indifferently to both this family and ${\mathfrak F} = \{ X :
\exists \tau \in \Pi \wedge X = {\mathfrak M}(\tau)\}$ as {\em 
the frame\/} for ${\mathfrak L}$ based on {\it D}.

A {\em denotation\/} for ${\mathfrak L}$ based on {\it D} is a quasi-function
$\phi$ whose domain is the set of constants of ${\mathfrak L}$, defined
such that $\phi (A^{\tau}) \in {\mathfrak M}_{\tau}$ for every 
$\tau \in \Pi$. So, in particular  $\phi (A^{e_{1}}) \in m$
and $\phi (A^{e_{2}}) \in M$. 

Based on these definitions, we may introduce the concept of
an {\it interpretation\/} for ${\mathfrak L}$ based on $D$ as an ordered pair
${\mathfrak A} = \langle ({\mathfrak M}_{\tau})_{\tau \in
\Pi}, \phi \rangle$, where $({\mathfrak M}_{\tau})_{\tau \in
\Pi}$ is a frame for ${\mathfrak L}$ (based on $D$) and $\phi$ a denotation 
as above. The interpretation is {\em principal\/} if the frame
is standard. It can be shown that the defined predicate of equality
 (Definition 1.1) is interpreted in the quasi-set $\Delta_{\equiv}(\tau)$, 
the `pseudo-diagonal' of ${\mathfrak M}_{\tau}$, 
namely, the qset whose elements are pairs of indistinguishable objects
(see \cite{dckra97i}).

A {\it valuation\/} for ${\mathfrak L}$ (based on $D$) is a
quasi-function $\psi$ whose domain is the collection of terms of
${\mathfrak L}$ and in such a way that $\psi$ is the extension of the
denotation quasi-function $\phi$ to the whole set of terms of
${\mathfrak L}$. In other words, $\psi$ is defined as follows: 

\begin{enumerate}
\item $\psi (A^{\tau}) \equiv \phi (A^{\tau})$ for every constant 
$A^{\tau}$. That is, the images of $A^{\tau}$ by $\phi$ and by $\psi$ are
indistinguishable. 
\item $\psi (x^{e_{1}}) \in m$
\item $\psi (X^{e_{2}}) \in M$
\item $\psi (X^{\tau}) \in {\mathfrak M}_{\tau}$ for $\tau \not= e_{1}, e_{2}$.
\end{enumerate}

We introduce the concept of a formula $A$ being
 {\it satisfiable\/} with respect to the
interpretation ${\mathfrak A}$  (in symbols: ${\mathfrak A}, \psi \models A$)  
in accordance with the following clauses:

\begin{enumerate}
\item  ${\mathfrak A}, \psi \models U^{\tau}(X^{\tau}_{1}, \ldots, 
X^{\tau}_{n})$ iff $\langle \psi (X^{\tau}_{1}), \ldots, \psi
(X^{\tau}_{i}) \rangle \in \psi (U^{\tau})$, where $U^{\tau}$ is
a term of type $\langle \tau_{1}, \ldots, \tau_{n} \rangle$ and
$X^{\tau}_{n}$ are terms of type $\tau_{i}$ ($i = 1, \ldots, n$).
\item  The satisfaction clauses for $\neg$, $\to$ and
$\forall$ are introduced as usual.
\end{enumerate}

A formula $A$ is {\it true\/} with respect to the
interpretation ${\mathfrak A}$ iff ${\mathfrak A}, \psi \models A$ for
every valuation $\psi$ with respect to ${\mathfrak A}$. 
 An interpretation ${\mathfrak A}$ is {\it
normal\/} iff every instance of the axioms of ${\mathfrak L}_{op}$ is true
in ${\mathfrak A}$, as are all instances of 
extensionality, separation, infinite and choice.
In what follows we will consider only appropriate interpretations.

A normal interpretation which is not principal is a
{\it secondary\/} interpretation. A formula
$A$ is {\it valid\/} iff it is true with respect to all principal
interpretations, and it is {\it satisfiable\/} if there exists a principal
interpretation ${\mathfrak A}$ and a valuation $\psi$ such that
${\mathfrak A}, \psi \models A$. The formula is {\it secondarily valid\/}
if it is true with respect to all normal interpretations, and it is
{\it secondarily satisfiable\/} if is is true with respect to some
normal interpretation. 

Then, by adapting the proofs presented in \cite{chu56}, we can state without
difficulty the following results: (1) $A$ is valid iff $\neg
A$ is not satisfiable; (2) $A$ is secondarily valid iff $\neg A$
is not secondarily satisfiable; (3) $A$ is satisfiable iff $\neg
A$ is not valid; (4) $A$ is secondarily satisfiable iff $\neg A$
is not secondarily valid and (5) $A$ is valid (respect.
secondarily valid) with respect to a normal interpretation iff
its universal closure is valid (respect.  secondarily valid)
with respect to this interpretation (see also \cite{coskra94}). 

By a {\it model\/} of a set $\Gamma$ of formulas of ${\mathfrak  L}$ 
we understand a normal interpretation ${\mathfrak A}$ such
that ${\mathfrak A}, \psi \models A$ for every formula $A \in
\Gamma$. If ${\mathfrak A}$ is a principal interpretation, we talk of {\em
principal models\/}, or of {\em secondary models\/} if ${\mathfrak A}$
is a secondary interpretation. The following terminology will be
used below: $\Gamma \models A$ means that $A$ holds in every
model of $\Gamma$, and $\models$ $A$ means that $A$ is
secondarily valid.

The proofs of the theorems below are simple adaptations from those
of usual higher-order logic \cite{chu56}, and we don't think they must
be repeated here. A particular case of similar results involving
quasi-set semantics was presented with more details in \cite{dckra97i}.

\begin{theorem}[Soundness]
All theorems of ${\mathfrak L}_{op}$ are secondarily valid.
\end{theorem}

In other words, $\vdash A$ implies $\models A$; it is not difficult to
generalize this result: $\Gamma \vdash A$ implies $\Gamma
\models A$.

\begin{theorem}[Lindenbaum] Every consistent set $\Gamma$ of closed
formulas of $\mathfrak L$ can be extended to a maximal consistent
class $\overline{\Gamma}$ of closed formulas of $\mathfrak L$.\footnote{The
concepts mentioned here are like the standard ones.}
\end{theorem}

Furthemore, there is the following important result:

\begin{theorem}
If $A$ is a closed formula of $\mathfrak L$ which is not a theorem,
then there exists a normal interpretation whose domains ${\mathfrak
M}_{\tau}$ are denumerably infinite, with respect to which
$\neg A$ is valid.
\end{theorem}

Then, based on these theorems,
we can state the (weak) completeness theorem for our logic, whose
proof can be adapted either from \cite{chu56} or \cite{dckra97i}.

\begin{theorem}[Completeness]  Every formula of ${\mathfrak L}_{op}$
which is secondarily valid is a theorem.
\end{theorem}

That is, $\models A$ implies $\vdash A$. In general, if $\Gamma$
is a set of closed formulas of ${\mathfrak L}$ which is not
inconsistent, then $\Gamma \models A$ implies $\Gamma \vdash A$,
that is, if $A$ holds in every model of $\Gamma$, then $A$ is
derivable from the formulas of $\Gamma$.


\section{Veiled Sets}
Having sketched the main features of the generalized
quasi-set semantics for our logic, let us turn to a consideration of  the
opaque predicates and see more carefully how they were interpreted in the
framework of the previous section.

An opaque relation, according to our previous definiton, is a term
of type $\langle \tau \rangle$ where $\tau \in \Pi$ obtained
recursively from the basic type $e_{1}$. Intuitively, an opaque
predicate is an unary opaque relation of type $\tau = \langle e_{1} \rangle$.
Semantically, to an opaque predicate is associated  a 
subquasi-set of the pure quasi-set $m$.
In other words, the extension of such a predicate is a
collection of objects for which there is no sense in saying that
they are equal or distinct, and this is in conformity with the
above mentioned examples from quantum mechanics. So, the semantics of
our logic agrees with its syntatic aspects. 

Such pure quasi-sets, let us remark, seem to be concealed by a
kind of veil, since although they have a well-determined
neighbourhood (the membership relation has a standard
behaviour),\footnote{That is, their characteristic function is a
(quasi)function from the quasi-set in $\{0, 1\}$, as in usual
set-theory, what this intuitively means that, for every $x$, 
 $x$ belongs or  does not belong to the quasi-set. So, the
present case is distinct from that involving fuzzy sets or
{\it quasets\/} (concerning the latter, see \cite{dalgiukra95}.} we
definitely cannot distinguish between their elements.
Furthermore, the `unobservability permutation theorem' mentioned
above can be understood as saying that if an element of one of
these quasi-sets is exchanged by an indistinguishable one, all
happens (to us, who are behind the veil) as if nothing had occurred
at all.\footnote{The case of the unobservability of permutations
in quantum mechanics is `didactically' mentioned by Roger Penrose
in his {\it The emperor's new mind\/}: ``according to the modern
theory [quantum mechanics], if a particle of a person's body
were exchanged with a similar particle in one of the bricks of
his house then nothing would have happened at all''
\cite[p. 360]{pen89}.}   Such quasi-sets could be
called {\it veiled sets\/}, and are the `natural' 
extensions of opaque predicates. Furthermore, in accordance with
quantum mechanics, indistinguishable $m$-atoms are not like objects 
(individuals) which
merely cannot be identified; there is a strong `ontic' indeterminacy 
among them \cite{frekra95}. 

\vspace{3mm}
Before ending the paper, let us comment briefly about a
peculiarity of  our logic ${\mathfrak L}_{op}$. We have
choosen such a logic as `the' logic of opaque predicates on the grounds
of formalizing the intuitive idea that opaque
predicates should be characterized as describing properties for 
which it is not possible to distinguish between the elements
that have the property ascribed by the predicate. 
 Of course it would be incorrect to
suggest that ${\mathfrak L}_{op}$ is the only logic of such
predicates. Notwithstanding this, someone could ask us why we
didn't  use classical logic instead of ${\mathfrak L}_{op}$ but
in such a way  that its semantics is represented in quasi-set
theory.  Then, he/she could say, it should be sufficient to
interpret some predicates of the language as veiled sets and the
idea of opaque predicates could be achieved. This of course is
an interesting idea which could simplify much of the above discussion.
 But, although we agree with the convenience of using
classical logic when possible, in this case we would have no
syntactical means of distinguishing opaque predicates from 
other predicates: only a  `semantical' distinction would be provided by
using veiled sets as the extensions of some predicates,
 while to the remaining ones standard sets should be used instead. 
But we think that we could try  to obtain a logical system which provides
also a syntactical distinction between the predicates, and our logic
may be viewed as an attempt in this direction.
In fact, our system provides not only a distinction among predicates of
its language, but by
postulating that the concept of identity is meaningless for certain
entities, it is still in accordance with the intuitive idea
of characterizing opaque predicates. So, returning to
Peirce's characterisation,  perhaps we can say the
following: A proposition is opaque when there are possible
states of things concerning which it is intrinsically uncertain
whether, had they been contemplated by the speaker, he would
have regarded them as excluded or allowed by the proposition. By
intrinsically uncertain we mean not uncertain in consequence of
any ignorance of the interpreter, or any indeterminacy in the
speaker's habits of language but because of a fundamental
ontological indeterminacy with regard to the objects denoted.


\begin{thebibliography}{99}
\bibitem{bla66} Black, M., {\it Language and philosophy\/},
Ithaca, NY, Cornell Univ. Press, 1966. 

\bibitem{bor43}Born, M., {\it Experiment and theory in physics\/},
Cambridge, Cambridge un. Press, 1943.

\bibitem{can58} Cantor, G., {\it Contributions to the founding of the theory
of transfinite numbers\/},  New York, Dover Pu., 1955.

\bibitem{chu56} Church, A., {\it Introduction to mathematical logic\/},
Vol. 1, Princeton, Princeton Un. Press, 1956.

\bibitem{cosfre90} da Costa, N. C. A., and French, S., `The model theoretical
approach in the philosophy of science', {\it Philosophy of Science\/}
{\bf 57}, 1990, 248--265. 

\bibitem{coskra94} da Costa, N. C. A. and Krause,
D., `Schr\"odinger logics', {\em Studia Logica\/} {\bf 53}, 1994, 533--550.

\bibitem{dckra97i} da Costa, N. C. A. and Krause,
D., `An intensional Schr\"odinger logic', 
{\it Notre Dame J. of Formal Logic\/} {\bf 38} (2), 1997, 179--194.

\bibitem{coskra95f} da Costa, N. C. A. and Krause,
D., `Set-theoretical models for quantum systems', 
forthcoming
in Dalla Chiara, M. L., Giuntini, R. and Laudisa, F. (eds),
{\it Philosophy of Science in Florence, 1995\/}, Kluwer Ac. Press.
 Abstract in the {\it Volume of Abstracts\/},
Xth International Congress of Logic, Methodology and Philosophy of Science,
Florence, Aug., 19--25, 1995, p. 470. 

\bibitem{dal85} Dalla Chiara, M. L., `Some foundational
problems in mathematics suggested by physics', {\em Synthese\/}
{\bf 62}, 1985, 303-315.

\bibitem{dal87} Dalla Chiara, M. L.  `An approach to intensional 
semantics', {\em Synthese\/} {\bf 73}, 1987, 479--496.

\bibitem{daltor93} Dalla Chiara, 
M. L. and Toraldo di Francia, G., 
`Individuals, kinds and names in physics',  in Corsi, G. et al. (eds.),
 {\em Bridging the gap: philosophy, mathematics,
physics\/}, Dordrecht, Kluwer Ac. Press, 1992, 261--283.

\bibitem{daltor93b}
Dalla Chiara M. L. and Toraldo di Francia, G., `Identity
questions from quantum theory', in Gavroglu et. al. (eds.), {\it Physics,
philosophy and the scientific community\/}, Dordrecht, Kluwer, 1995,
39--46.

\bibitem{dalgiukra95} Dalla Chiara, M. L., Giuntini, R. and Krause, D.,
`Quasiset theories for microobjects: a comparision', forthcoming in
E. Castellani (ed.),
{\it Interpreting bodies: classical and quantum objects in modern
physics\/}, Princeton Un. Press.


\bibitem{fre89} 
French, S., `Identity and individuality in classical and
quantum physics', {\em Australasian Journal of Philosophy\/}
{\bf 67}, 1989,  432--446.

\bibitem{frered88} French, S. and Redhead, M., 
`Quantum physics and the identity of indiscernibles', {\em
British Journal for the Philosophy of Science\/} {\bf 39} (1988), 233-246.

\bibitem{frekra95} 
French, S. and Krause, D., `Va\-gue iden\-tity and 
quantum \- non-indi\-vidual\-ity',  {\em Analysis\/} {\bf 55}
(1),  1995, 20--26.

\bibitem{frekra96} French, S. and Krause, D., `The logic of quanta',
forthcoming 
in Cao, T. L., (ed.) {\it Proceedings of the Boston
Colloquium for the Philosophy of Science 1996: A historical examination
and philosophical reflections on the foundations of quantum field theory\/}, 
Cambridge, Cambridge Un. Press.

\bibitem{frekramai95} French, S., Krause, D. and Maidens, A., `Quantum
vagueness', preprint, University of Leeds. 

\bibitem{hes70} Hesse, M., {\it Models and analogies in science\/},
Un. of Notre Dame Press, 1970 (1963).

\bibitem{hilack51} Hilbert, D. and Ackermann, W., {\it Principles of
mathematical logic\/}, New York, Chelsea, 1951. 

\bibitem{kra92} Krause, D., `On a quasi-set theory', 
{\em Notre Dame Journal of Formal Logic\/} {\bf 33} (3), 1992, 402--411. 

\bibitem{kra95a} Krause, D., `The theory of quasi-sets and
ZFC are equiconsistent', in Carnielli, W. A. and Pereira, L. C. (eds.),
{\it Logic, sets and information\/}, (Proceedings of the Xth Brazilian 
Conference on Mathematical Logic,
Itatiaia, 1993), UNICAMP, Col. CLE vol 14, 1995, 145--155. 

\bibitem{kra96} Krause, D., `Axioms for collections of indistinguishable
objects',  {\it Logique et Analyse\/} {\bf 153--154}, 1996, 69--93.

\bibitem{krafre95} Krause, D.  and French, S.,
`A formal framework
 for  quantum non-individuality', {\em Synthese\/} {\bf 102},
1995, 195--214. 

\bibitem{krasanvol98} Krause, D., Sant'Anna, A. S. and Volkov, A. G.,
`Quasi set theory for bosons and fermions', preprint, Federal University
of Paran\'a. 

\bibitem{low94} Lowe, J., `Vague identity and quantum indeterminacy',
 {\it Analysis\/} {\bf 54}, 1994, 110--114. 

\bibitem{mot94} Mott, P., `On the intuitionistic solution of the Sorites
Paradox', {\it Pacific Phil. Quarterly\/} {\bf 75}, 1994, 133--150.

\bibitem{par83} Parikh, R., `The problem of vague predicates',
in Cohen, R. S. and Wartofski, M. N. (eds.), {\it Language, logic and
method\/}, Dordrecht, Reidel, 1983, 241--261. 

\bibitem{pei02} Peirce, C. S., in Baldwin's {\it Dictionary of Philosophy and
Psychology\/} {\bf 2}, 1902, 748.  

\bibitem{pen89} Penrose, R., {\it The emperor's new mind\/},
Oxford, Oxford Un. Press, 1989.

\bibitem{pos63}
Post, H. `Individuality and physics',
{\em The Listener\/} {\bf 70}, 1963, 534-537. Reprinted in 
{\em Vedanta for East and West\/} {\bf 32}, 1973, 14-22.

\bibitem{put83} Putnam, H., `Vagueness and alternative logic', {\it
Erkenntnis\/} {\bf 19}, 1983, 297--314. 

\bibitem{ram78} Ramsey, F. P., {\it Foundations: essays in philosophy,
logic, mathematics and economics\/}, ed. by D. H. Mellor, London and Henley,
Routledge \& Kegan Paul, 1978 (1931). 

\bibitem{sch52} Schr\"odinger, E., {\em Science and humanism\/}, 
Cambridge Un. Press, Cambridge, 1952. 

\bibitem{sch57} Schr\"odinger, E., {\em Science theory 
and man\/}, Allen and Unwin, London, 1957.

\bibitem{tel95} Teller, P., {\em An interpretive introduction to
quantum field theory\/}, Princeton, Princeton Un. Press, 1995.

\bibitem{tertri89} Terricabras, J. -M. and Trillas, E., `Some remarks
on vague predicates', {\it Theoria -- Segunda \'Epoca\/} {\bf 10}, 1989,
1--12.

\bibitem{van91} van Fraassen, B., {\it Quantum mechanics: an empiricist
view\/}, Oxford, Clarendon Press, 1991. 

\bibitem{wri76} Wright, C., `Language-mastery and the Sorites paradox',
in Evans, G. and McDowell, J., {\it Essays in semantics\/}, Oxford,
Clarendon Press, 1976, 223--247. 

\end{thebibliography}
\end{document}